\documentclass[useAMS,usenatbib]{PoS}
\usepackage{graphicx}
\usepackage{subfigure}

\def\kms      {\ifmmode{\rm km\,s}^{-1} \else km\,s$^{-1}$\fi}
\def\mujybm{\ifmmode{\rm \mu Jy}\,{\rm beam}^{-1}\else${\rm \mu}$Jy\,beam$^{-1}$\fi}
\def\ltsim{\ifmmode\stackrel{<}{_{\sim}}\else$\stackrel{<}{_{\sim}}$\fi}
\def\gtsim{\ifmmode\stackrel{>}{_{\sim}}\else$\stackrel{>}{_{\sim}}$\fi}
\def\farcs{\hbox{$.\!\!^{\prime\prime}$}}
\def\farms{\hbox{$.\!\!^{\prime}$}}
\def\arcsec{\hbox{$^{\prime\prime}$}}

\def\hour{\hbox{$^{\rm h}$}}

\def\mum{$\mu$m}
\def\spitzer{{\it Spitzer}}
\def\q24{q$_{\rm 24}$}

\title{An evolution of the IR-Radio correlation?}

\ShortTitle{An evolution of the IR-Radio correlation?}

\author{\speaker{R. J. Beswick}\\
        Jodrell Bank Centre for Astrophysics, The University of Manchester\\
        E-mail: \email{Robert.Beswick@manchester.ac.uk}}

\author{T. W. B. Muxlow\\
        Jodrell Bank Centre for Astrophysics, The University of Manchester\\
        E-mail: \email{twbm@jb.man.ac.uk}}

\author{H. Thrall\\
        Jodrell Bank Centre for Astrophysics, The University of Manchester\\
        E-mail: \email{hthrall@jb.man.ac.uk}}

\author{A. M. S. Richards\\
        Jodrell Bank Centre for Astrophysics, The University of Manchester\\
        E-mail: \email{amsr@jb.man.ac.uk}}

\author{S. T. Garrington\\
        Jodrell Bank Centre for Astrophysics, The University of Manchester\\
        E-mail: \email{stg@jb.man.ac.uk}}

\abstract{Using extremely deep (rms $\sim$3.3$\mu$Jy/bm) 1.4\,GHz
             sub-arcsecond resolution MERLIN$+$VLA radio
             observations of a 8\farms5$\times$8\farms5 field
             centred upon the Hubble Deep Field North, in
             conjunction with {\it Spitzer} 24\,$\mu$m data we
             present an investigation of the radio-MIR correlation
             at very low flux densities.  By stacking individual
             sources within these data we are able to extend the
             MIR-radio correlation to the extremely faint
             ($\sim$microJy and even sub-microJy) radio source
             population. Tentatively we demonstrate a small
             deviation from the correlation for the faintest MIR
             sources. We suggest that this small observed change in
             the gradient of the correlation is the result of a
             suppression of the MIR emission in faint star-forming
             galaxies.  This deviation potentially has significant
             implications for using either the MIR or non-thermal
             radio emission as a star-formation tracer at low
             luminosities.
}

\FullConference{From planets to dark energy: the modern radio universe\\
		 October 1-5 2007\\
		 University of Manchester, Manchester, UK}

\begin{document}

\section{Introduction}

Radio and infrared emission from galaxies in both the nearby and distant Universe is 
thought to arise from processes related to star-formation, hence resulting in the 
correlation between these two observing bands. The infrared emission is produced from 
dust heated by photons from young stars and the radio emission predominately arises 
from synchrotron radiation produced by the acceleration of charged particles from 
supernovae explosions. It has however recently been suggested that at low flux 
density and luminosities there may be some deviation from the tight well-known 
radio-IR correlation seen for brighter galaxies \cite{bell03,boyle07}.

\cite{bell03} argue that while the IR emission from luminous galaxies will trace the 
majority of the star-formation in these sources, in low luminosity galaxies the IR 
emission will be less luminous than expected considering the rate of star-formation 
within the source (i.e. the IR emission will not fully trace the star-formation). In 
this scenario the reduced efficiency of IR production relative to the source 
star-formation rate (SFR) would be the result of inherently lower dust opacities in 
lower luminosity sources and consequently less efficient reprocessing of UV photons 
from hot young stars into IR emission. The simple consequence of this is that at 
lower luminosities the near linear radio-IR correlation L$_{\rm radio}\propto$L$_{\rm 
IR}^{\!\gamma}$, with $\gamma >1$ (e.g. \cite{cox88,price92}) will be deviated from. 
{\it Of course such an assertion is dependent upon the radio emission providing a 
reliable tracer of star-formation at low luminosities which may be equally invalid.}

Recently \cite{boyle07} have presented a statistical analysis of Australia 
Telescope Compact Array (ATCA) 20\,cm observations of the 24\,\mum\ sources 
within  the \spitzer\ Wide Field Survey 
(SWIRE). In this work \cite{boyle07} have co-added sensitive 
(rms$\sim$30\,$\mu$Jy) radio data at the locations of several thousand 24\,\mum\ 
sources. Using this method they have statistically detected the microJy radio 
counterparts of faint 24\,\mum\ sources. At low flux densities (S$_{\rm 24\, \mu 
m }=100\,\mu$Jy) they confirm the IR-radio correlation but find it to have a 
lower coefficient (S$_{\rm 1.4\,GHz}$\,=\,0.039\,S$_{\rm 24\,\mu m}$) than had 
previously been reported at higher flux densities. This coefficient is 
significantly different from results previously derived from detections of 
individual objects (e.g. \cite{appleton04}) and is speculated by 
\cite{boyle07} to be the result of a change in the slope of the radio-IR 
correlation at low flux densities.

In this work (which is descirbed in more detail in \cite{beswick06, beswick08})
we have utilised very deep, high resolution 20\,cm observations of the 
Hubble Deep Field North and surrounding area made using MERLIN and the VLA 
\cite{muxlow05} in combination with publicly available 24\,\mum\ \spitzer\ 
source catalogues from GOODS to study the MIR-Radio correlation for microjansky 
radio sources. This study extends the flux density limits of the radio-IR 
correlation by more than an order of magnitude for individual sources and 
overlaps the flux density regime studied using statistical stacking
methods by other authors.

 \begin{figure*}
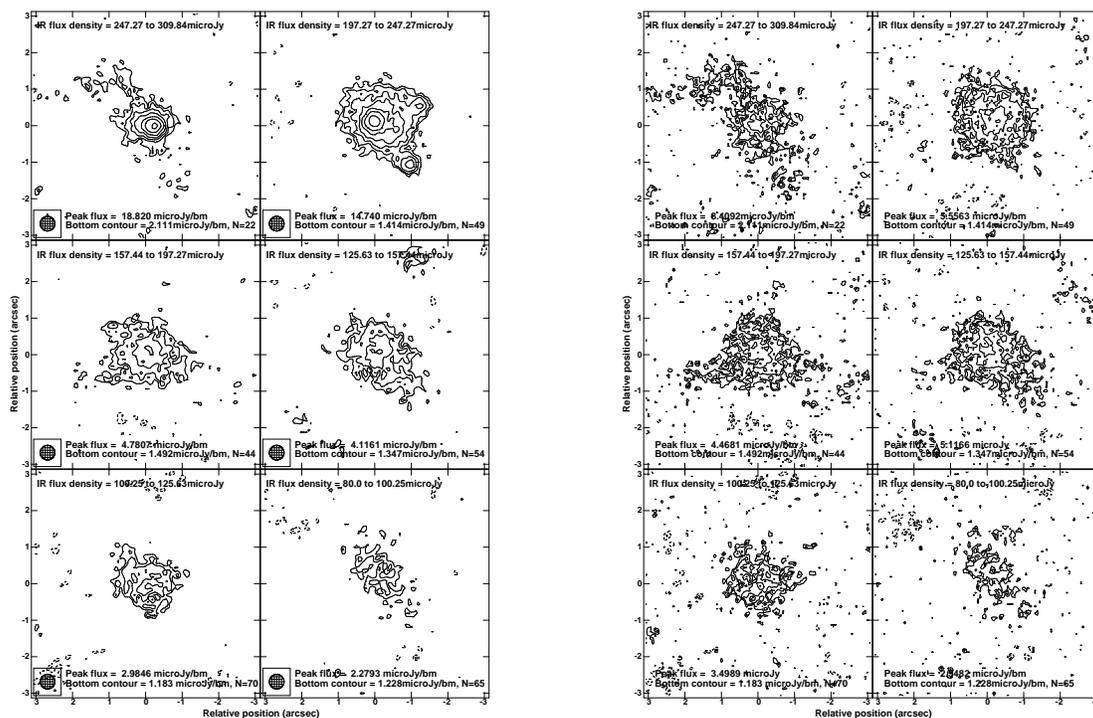

\begin{center}$
\begin{array}{cc}
\includegraphics[width=8cm]{Beswick-fig3a.ps} &
\includegraphics[width=8cm]{Beswick-fig3b.ps}
\end{array}$
\caption{Mean (left) and median (right) images of the 1.4\,GHz radio emission 
for all sources within the six faintest 24\,\mum\ flux density logarithmic bins plotted 
in Fig.\,1 in descending flux density order from top-left to bottom-right. The 
range of 24\,\mum\ flux density over which each image has been stacked is shown 
at the top of individual panels. Each image is contoured with levels of $-$2, 
$-$1.414, $-$1,1, 1.414, 2, 2.828, 4, 5.657, 8, 11.31, 16, 22.63 and 32 times 
3$\times$(3.3/$\sqrt {\rm N}$)\,$\mu$Jy\,bm$^{-1}$, where N equals the number of 
24\,\mum\ source positions averaged in the map. The peak flux density, lowest 
plotted contour and number of IR sources which have been averaged over (N) is 
listed at the bottom of each image panel.}
\label{maps}
\end{center}
 \end{figure*}

\section{Observations}

\subsection{Radio Data}

Extremely deep radio observations of the HDF-N region were made in 1996-97 at 
1.4\,GHz using both MERLIN and the VLA. These observations were initially 
presented in \cite{muxlow05}, \cite{richards98} and \cite{richards00}. The 
results from the combined 18 day MERLIN and 42\,hr VLA observations are described 
in detail in \cite{muxlow05}. The combined MERLIN$+$VLA image has an rms noise level of 
3.3\,$\mu$Jy per 0\farcs2 circular beam making it amongst the most sensitive, high-resolution radio images made to date.

\subsection{GOODS-N \spitzer\ 24\,$\mu$m observations}

As part of the GOODS enhanced data 
release\footnote{http://www.stsci.edu/science/goods/DataProducts/} (DR1+ February 
2005) a catalogue of \spitzer\ 24\,$\mu$m source positions and flux densities for 
the GOODS-N field were released. This source catalogue is 
limited to flux densities $>$80\,$\mu$Jy providing a highly complete and reliable 
sample. At the time of writing this 24\,$\mu$m source catalogue represents the 
most complete and accurate mid-infrared source list publicly available for the 
GOODS-N/HDF-N field.

All 24\,\mum\ sources which are detected optically in GOODS {\it HST} ACS images 
show an accurate astrometric alignment with their optical counterparts implying 
that the astrometry between these two data-sets and their subsequent catalogues 
is self-consistent. However, a comparison of the astrometric alignment of the 
positions of sources catalogued by GOODS derived from their {\it HST} ACS images 
\cite{richards07,beswick08} shows there to be a systematic offset in 
declination of $-$0\farcs342 from the radio reference frame.  This linear 
declination correction, although small relative to the \spitzer\ resolution at 
24\,\mum\, is significant when compared to these high resolution radio data. This 
linear correction has been applied to the \spitzer\ source positions prior to all 
comparisons between the two data sets.

\begin{figure}[ht!]
\begin{center}$
\begin{array}{cc}
\includegraphics[width=7.5cm]{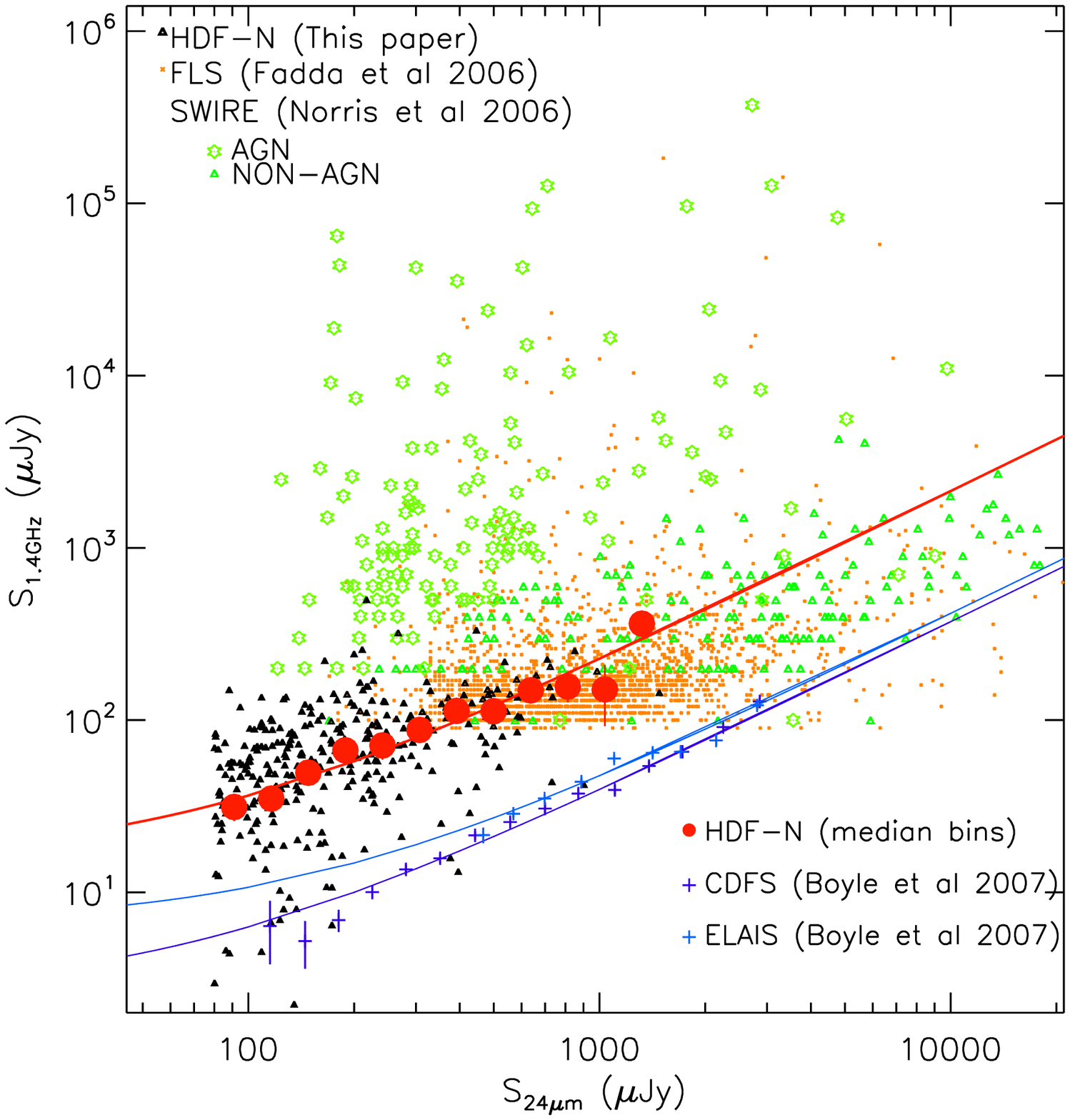} &
\includegraphics[width=7.5cm]{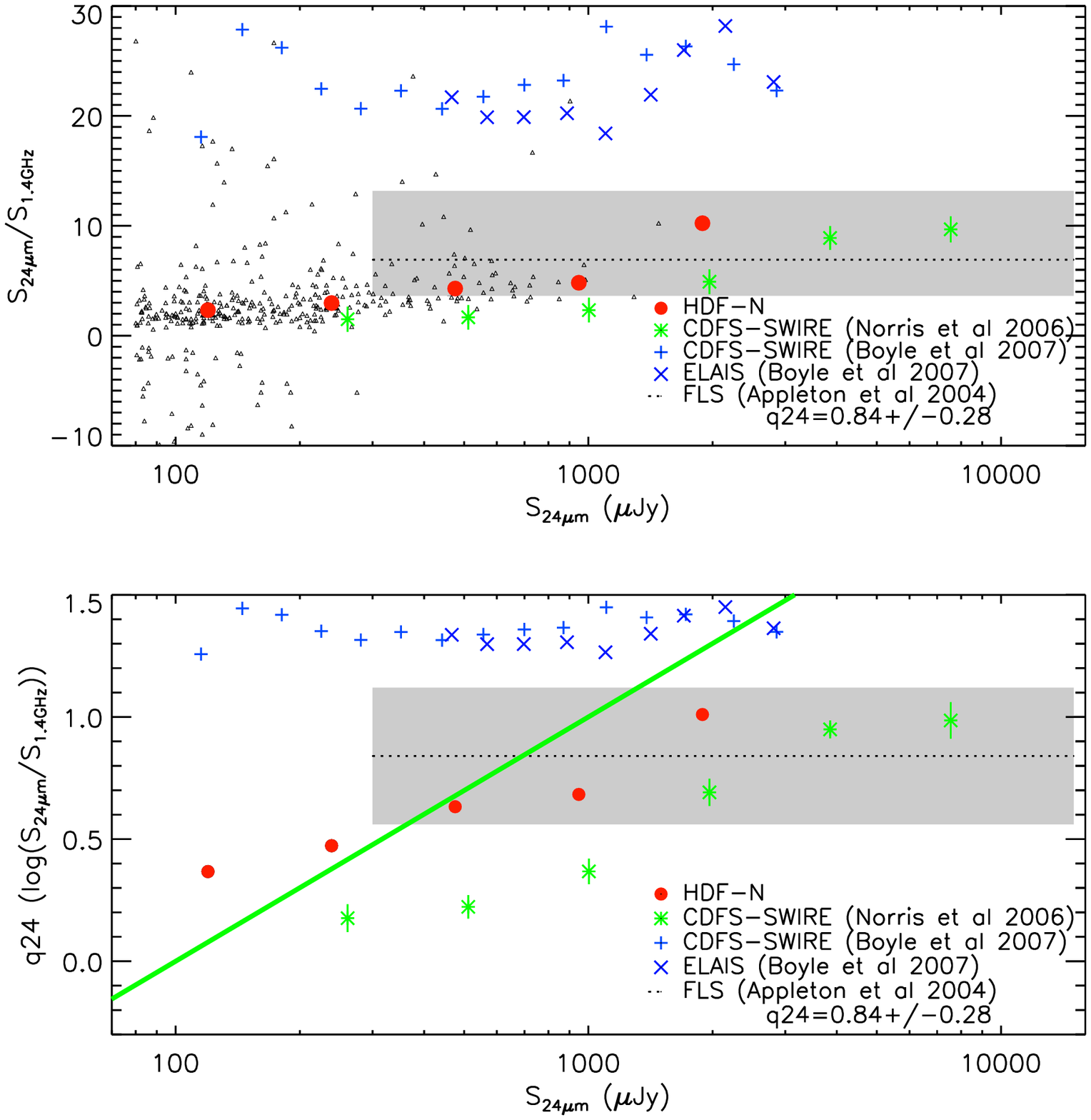}
\end{array}$
\caption{{\bf Left:} Radio 1.4\,GHz versus the 24\,\mum\ flux density. Sources from the 
8\farms5$\times$8\farms5 HDF-N field are plotted individually (small black 
triangles). The median radio flux density logarithmically binned by 24\,\mum\ 
flux densities are plotted as large filled red circles. The solid red line represents the best-fit line to 
the binned HDF-N data alone. Sources within the CDFS-SWIRE field detected at both 
24\,\mum\ and 1.4\,GHz from Norris et al. \cite{norris06} are plotted as either 
green open stars (identified as AGN) or green triangles (not identified as AGN). 
All sources detected at both 1.4\,GHz and 24\,\mum\ in the \spitzer\ First Look 
Survey (FLS) with source position separations of $<$1\farcs5 are plotted in 
orange \cite{fadda06}.  Note the quantisation of the SWIRE and FLS points, in 
this and subsequent plots is a result of the accuracy of the flux densities 
tabulated in the literature. The blue pluses and fitted lines in the low portion 
of the plot show the IR-radio correlation derived from stacking radio emission at the positions of 24\,\mum\ 
sources in the CDFS and ELAIS field by Boyle et al. \cite{boyle07}.\newline
{\bf Right:} In the upper panel the flux density ratio (${\rm S_{24\,\mu m}}/{\rm 
S_{1.4\,GHz}}$) versus 24\,\mum\ flux density is shown. The individual 
sources from the HDF-N field are plotted as small black triangles, the median values of these 
HDF-N source binned as a function of S$_{24\,\mu m}$ as filled circles, green 
stars show the median binned values for sources listed as non-AGN within the 
CDFS-SWIRE sample of Norris {\it et al.} \cite{norris06} and the blue `crosses' and `pluses' show the flux density ratios derived from the stacking analysis of the 
SWIRE-CDFS and ELAIS fields respectively from Boyle {\it et al.} \cite{boyle07}. The overlaid 
black dotted line is the mean value for (${\rm S_{24\,\mu m}}/{\rm 
S_{1.4\,GHz}}$) derived by Appleton {\it et al.} \cite{appleton04}. This line is equivalent 
to q$_{24}$=0.84$\pm$0.28 with the gray filled box representing the area enclosed 
by these errors and the flux density range investigated by Appleton {\it et al.}. The values of q$_{24}$ against 24\,\mum\ flux density are plotted in the 
lower panel.  The symbols within this plot are identical to those in the upper 
panel, individual HDF-N sources are not included for clarity. The additional 
diagonal solid green line represents a line of constant radio flux density of 
100\,$\mu$Jy, the lowest flux density of sources in the SWIRE sample plotted here 
(sources above this line are excluded by this limit from the SWIRE sample). This 
flux density limit will significantly effect the values of the binned SWIRE data 
points (green crosses) negatively biasing the four lowest flux density bins. This 
bias only effects the SWIRE sample. }
\label{F20vsF24log}       
\end{center}
\end{figure}

\section{Results \& Discussion}

Using these two highly sensitive data sets the radio emission from a
sample of faint 24\,\mum\ galaxies has been invesitigated. Of the 377
\spitzer\ sources within the radio field 303 were found to have total
radio flux denisties in excess of 3 times the local rms in our deep
radio imaging. Many of these sources, however, have peak flux
densities comparable or below the point source detection threshold of
these radio data. 

By using both statistically stacked images and
radio flux density measurements at the positions of individual
\spitzer\ 24\,\mum\ sources we have investigated the radio structures and
flux densities of faint IR sources. In figure \ref{maps} the
statistically averaged radio emission from faint 24\,\mum\ galaxies in
the GOODS-N field are shown. Each of 
these images is the statistically combined radio emission from the location of 
several tens of \spitzer\ 24\,\mum\ sources and has been contoured at multiples 
of three times 3.3\,$\mu$Jy\,beam$^{-1}$ 
divided by $\sqrt{\rm N}$ where N equals the number of \spitzer\ source positions 
that have been stacked together. As can be seen in these images the off-source 
noise levels achieved approaches the value expected when co-adding multiple images 
with near-Gaussian noise properties. The co-added image rms achieved in the faintest 24\,\mum\ flux density bin (80.0 to 100.25\,$\mu$Jy) is 
0.45 and 0.56\,$\mu$Jy\,beam$^{-1}$ in the mean and median 
co-added images respectively. 

The Gaussian fitted sizes of the radio emission in the stacked images 
(Fig\,\ref{maps}) provide an upper limit on the average size of the radio 
counterparts of these faint IR sources. The largest angular sizes of the radio 
emission in the median stacked images created from this sample range between 
1\farcs4 and 2\arcsec. This is approximately equivalent to a linear size of 
10\,kpc at redshifts beyond 1. These upper limits on the radio source sizes are 
consistent with radio emission on galactic and sub-galactic scales and
originating within kpc-scale starburst systems. 

By combining the \spitzer\ 24\,\mum\  flux densities and the extracted
1.4\,GHz radio flux densities for these sources it is possible to
begin to characterise the radio emission of the faintest IR
galaxies. Shown in the left-hand portion of
figure\,\ref{F20vsF24log} is 1.4\,GHz total flux density of sources
within the 8\farms5$\times$8\farms5 MERLIN$+$VLA field plotted against
their 24\,\mum\ flux density. As a direct comparison also overlaid on this
diagram are source flux densities from various other deep
multi-wavelength observational campaigns of different regions of
sky. As can be seen these new HDF-N/GOODS-N data significantly extend the observed
IR-radio correlation for galaxies between these two wavelengths.   

Binning and re-plotting these data in terms of S$_{\rm 24\,\mu m }$/S$_{\rm
1.4\,GHz}$  and q$_{24}$ (log(S$_{\rm 24\,\mu m}$/S$_{\rm
1.4\,GHz}$) (see figure\ref{F20vsF24log} right) clearly show the slope
and tightness of this correlation. However, it can be seen from
figure\ref{F20vsF24log} that these data  appear to show that this
correlation begins to deviate for the faintest sources. This deviation
is small (and tentative) but implies that the faintest galaxies are
under-luminous at MIR wavelengths relative to their observed radio
wavelengths. This deviate, whilst small, has potentially direct implications on
the use of either of these bands to quantify star-formation rates in
faint galaxies. A more complete discussion of these results and their
implications can be found in \cite{beswick06,beswick08}.

\section{Conclusion}
Using one of the deepest high-resolution 1.4\,GHz observations made to date, 
in conjunction with deep 24\,\mum\ \spitzer\ source catalogues from GOODS, we have 
investigated the microJy radio counterparts of faint MIR sources. These 
observations confirm that the microJy radio 
source population follow the MIR-radio correlation and extend this correlation by several orders of magnitude to 
very low flux densities and luminosities, and out to moderate redshifts. This extension 
of the MIR-radio correlation confirms that the majority of these
extremely faint radio and 
24\,\mum\ sources are predominantly powered by star-formation with little AGN 
contamination.

Statistically stacking the radio emission from many tens of faint
24\,\mum\ sources has been used to characterise the size and nature of the radio 
emission from very faint IR galaxies well below the nominal radio sensitivity of 
these data. Using these methods the MIR-radio correlation has been further 
extended and a tentative deviation in this correlation at very low 
24\,\mum\ flux densities has been identified.

\end{document}